\documentclass[a4paper,10pt]{article}
\usepackage{graphicx}

\begin{document}

\title{The open cluster Berkeley 53\thanks{This work is based on observations made with the 2-m
telescope of the Rozhen National Astronomical Observatory, which is
operated by the Institute of Astronomy, Bulgarian Academy of Sciences.}}
\author{G. Maciejewski\footnote{gm@astri.uni.torun.pl}\\
\small Centrum Astronomii UMK, ul. Gagarina 11, 87-100 Toru\'n\\ 
\small Astrophysikalisches Institut und Universit\"ats-Sternwarte, Schillerg\"asschen 2-3, D-07745 Jena, Germany\\
B. Mihov, Ts. Georgiev \\
\small Institute of Astronomy, Bulgarian Academy of Sciences, 72 Tsarigradsko Chausse Blvd., 1784 Sofia, Bulgaria}
\maketitle

\textit{Abstract}: We present a photometric study of the neglected open cluster Berkeley 53. We derived its fundamental parameters, such as the age, the interstellar reddening, and the distance from the Sun, based on $BV$ photometry combined with near-infrared $JHK_{\rm{S}}$ data. The structure and the mass function of the cluster were also studied and the total number of members and the total mass were estimated. The cluster was found to be a rich and massive stellar system, located in the Perseus Arm of the Milky Way, $3.1\pm0.1$ kpc from the Sun. Its age exceeds 1 Gy but it seems to be very young in the context of its dynamical evolution. The analysis of the two-color diagrams and color-magnitude diagrams indicates that the cluster is significantly reddened. However, both methods resulted in different values of $E(B-V)$, i.e. $1.21\pm0.04$ and $1.52\pm0.01$, respectively. This discrepancy suggests the presence of an abnormal interstellar extinction law toward the cluster.
 

\section{Introduction}

The open cluster Berkeley 53 (C 2055+508) was discovered by Setteducati \& Weaver (1960). Ruprecht (1966) classified it as a poor, concentrated open cluster of Trumpler type II3p. According to the \textit{New catalogue of optically visible open clusters and candidates} by Dias et al. (2002), the cluster's apparent diameter is $12'$ and Trumpler type is III2m. No dedicated studies of this object have been performed to date. 

Berkeley 53 consists of stars fainter than $V=18$ mag and is located in the vicinity ($3.\!'1$ from the cluster center) of the bright ($V=6.6$ mag) foreground star HD 199578. This makes photometric observations of the cluster difficult.

In this paper we present a photometric study of Berkeley 53 resulting in the determination of its basic parameters, such as the age, the interstellar reddening, and the distance from the Sun. The structure and the mass function of the cluster are also studied and the total number of members and the total mass are estimated.

\section{Observation and data reduction}

Observations were performed with the 2/16 m Ritchey-Chr\'etien telescope of the Rozhen National Astronomical Observatory (NAO, Bulgaria), operated by the Institute of Astronomy, Bulgarian Academy of Sciences. The instrument was used in a direct imaging mode and was equipped with a Princeton Instruments VersArray:1300B CCD camera mounted in the Ritchey-Chr\'etien focus. The field of view was $5.\!'8 \times 5.\!'6$ with a scale of $0.26$ arcsec per pixel. Observations were carried out on August 19, 2007. Two exposures in the $B$ filter and three frames in the $V$ band were acquired. The exposure time was 600 s and 300 s, respectively. Four exposures in the $U$ filter with 900 s exposure time were also acquired but they occurred to be not deep enough to detect the cluster stars. 

\begin{figure}
\centering
\includegraphics[width=8.3cm]{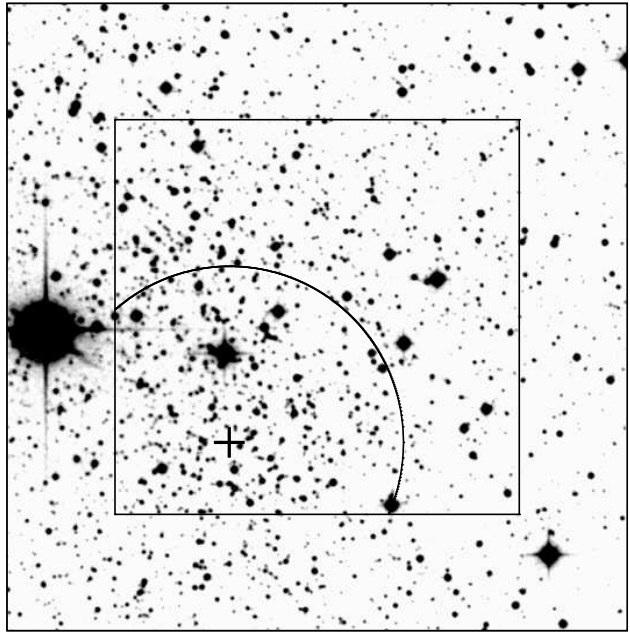}
\caption{A $ 9'$ $\times$ $9'$ field around Berkeley~53. The $5.\!'8$ $\times$ $5.\!'6$ field of view of the telescope is marked with a square frame. The redetermined center of the cluster is marked with a cross. The core radius boundary is also sketched. The bright star at the left edge of the image is HD 199578. The image was taken from the Digitized Sky Surveys (DSS). North is up, East to the left.}
\label{rys1}
\end{figure}

The field around Berkeley~53 is presented in Fig.~\ref{rys1} where the fragment of the sky covered by our observations is also sketched. The telescope was not pointed on the cluster center due to the nearby bright star HD~199578. The $B$ band exposures were affected by a narrow strip of its reflected light passing horizontally through the center of frames.

CCD frames were processed using a standard  procedure that included subtraction of bias frame, flat-fielding with twilight flats, aperture and point spread function (PSF) photometry, transformation to the standard system, and astrometric calibration. The aperture photometry and astrometric calibration were performed with the software pipeline developed for the Semi-Automatic Variability Search sky survey (Niedzielski et al. 2003). Exposures in a given band were averaged. The PSF magnitudes were obtained with IRAF\footnote{IRAF is distributed by the National Optical Astronomy Observatories, which are operated by the Association of Universities for Research in Astronomy, Inc., under cooperative agreement with the National Science Foundation.} package DAOPHOT. A second-order variable PSF was used to compensate for variability of stellar profiles across frames. Aperture corrections were determined using aperture photometry of 13 isolated stars that were used while building the PSF profile.

The raw instrumental magnitudes were corrected for the scattered light effect (Markov 2008). The following correction formulae were obtained by us:
\begin{equation}
 \Delta b = (0.122\pm0.006)\,d^2 \, , \;
\end{equation}
\begin{equation}
 \Delta v = (0.106\pm0.004)\,d^2 \, , \;
\end{equation}
where $\Delta b$, $\Delta v$ are the additive magnitude corrections in the corresponding filters and $d$ is the distance of a star from the CCD matrix center normalized by the detector half-size. 

The calibration coefficients that transform instrumental magnitudes into standard ones were determined using 16 stars from the Messier~92 field (Majewski et al.~1994, see their Table 1) and 40 stars from the field of NGC~7790 (Odewahn et al. 1992). The magnitude range was between $13.144$ and $18.315$ mag in $V$ and the $(B-V)$ color index coverage was in the range between $-0.111$ and $1.918$ mag. The following equations were derived:
\begin{equation}
 b - B = (0.58\pm0.03) X - (0.17\pm0.01)(B-V)-23.06 \, , \;
\end{equation}
\begin{equation}
 v - V = (0.33\pm0.02) X - (0.16\pm0.01)(B-V)-23.67\, , \;
\end{equation}
where $b$, $v$ are the corrected instrumental magnitudes, $B$, $V$ are the magnitudes in the standard system, and $X$ is the airmass. 

The final list of stars contains equatorial coordinates, $V$ magnitudes, and $(B-V)$ color indices. It is available in electronic form at the survey web site\footnote{http://www.astri.uni.torun.pl/\~{}gm/OCS} and the WEBDA\footnote{http://www.univie.ac.at/webda/} database (Mermilliod 1996).  

The optical data were complemented with near-infrared $JHK_{\mathrm{S}}$ photometry extracted from the 2-Micron All Sky Survey (2MASS, Strutskie et al.~2006). The extraction radius was set to $30'$ around the cluster center (see Sect. 3.1). Optical photometry was combined with near-infrared data to perform a comprehensive photometric study of the cluster (e.g. Maciejewski 2008).

\section{Data analysis and results}

\subsection{Cluster structure}

\begin{figure}
\centering
\includegraphics[width=7cm]{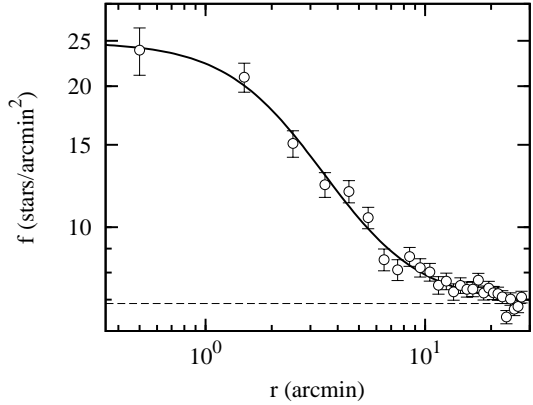}
\caption{The radial density profile. The solid line denotes the fitted density distribution while the dashed one marks the level of the stellar background density.}
\label{rys2}
\end{figure}

The cluster structure was investigated with the radial density profile (RDP). For this purpose only the 2MASS data set was used for constructing RDP due to unlimited field of view. We started with redetermining the cluster center. The algorithm was adopted from Maciejewski \& Niedzielski (2007) and started from a tentative position determined by eye. Two perpendicular stripes were cut along declination and right ascension starting from the approximate cluster center and the histogram of star counts was built along each stripe. The bin with the maximum value in both coordinates was taken as a new cluster center what resulted in $\mathrm{RA}=20^{\rm{h}}55^{\rm m}56^{\rm s}$ and $\mathrm{DEC}=+51^{\circ}02.\!'8$ for epoch J2000.0 ($l=90.\!\!^{\circ}29$, $b=3.\!\!^{\circ}75$). As one can see in Fig.~\ref{rys1}, these values do not point to the center of the field of view. Next, the profiles were constructed by counting stars inside concentric rings of width $1'$, centered at the redetermined cluster center. The density uncertainty in each ring was estimated assuming Poisson statistics. The RDP is plotted in Fig.~\ref{rys2}. 

To parametrize the density distribution, a two-parameter King (1966) density profile was fitted with the least-squares method in which the uncertainties were used as weights. We derived the core radius (the distance where the stellar density drops to half its maximum value) $r_{\mathrm{c}}=2.5\pm0.1$ arcmin, the central density $f_{0}=18.0\pm0.5$ stars/arcmin$^{2}$, and the background density level $f_{\mathrm{bg}}=6.9\pm0.1$ stars/arcmin$^{2}$. The fitted profile is sketched with a solid line in Fig.~2 while the dashed line marks the level of $f_{\mathrm{bg}}$. The cluster limiting radius $r_{\mathrm{lim}}$ was roughly estimated by eye-inspection in the RDP. The stellar density excess is visible up to at least $11'$~-- a value almost two times greater than the literature one.

The RDP allowed us to estimate number of observed stars belonging to the cluster to be $\sim$900. That suggests that Berkeley~53 is a very rich and massive stellar system.

\subsection{Two-color diagrams}

\begin{figure}
\centering
\includegraphics[width=10.3cm]{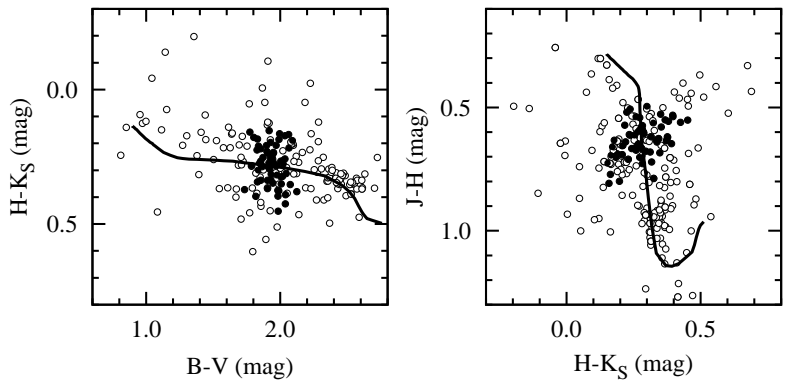}
\caption{$(H-K_{\mathrm{S}})$ $vs.$ $(B-V)$ and $(J-H)$ $vs.$ $(H-K_{\mathrm{S}})$ diagrams constructed for the stars located within the core radius from the cluster center. The stars that form the main sequence of the cluster are marked with filled circles. The main sequence, fitted by shifting along the reddening vector, is sketched with a continuous line in each diagram.}
\label{rys3}
\end{figure}

$(H-K_{\mathrm{S}})$ $vs.$ $(B-V)$ and $(J-H)$ $vs.$ $(H-K_{\mathrm{S}})$ two color diagrams (TCDs) were constructed to estimate the interstellar reddening toward the cluster. Only stars located within the core radius from the cluster center were considered to minimize the influence of the background star contamination. Among them only stars forming the main sequence of the cluster were taken for the further analysis. They are marked with filled circles in Fig.~\ref{rys3} where the TCDs are plotted.

The theoretical main sequence was extracted from the Padova isochrones for solar metallicity $Z=0.019$ (Giraldi at al. 2002). The reddenings $E(B-V)=1.18$ and $E(H-K_{\mathrm{S}})=0.26$ were obtained by shifting the main sequence (a continuous line in Fig. 3) along the reddening vectors whose normal slopes were calculated assuming the universal interstellar extinction law by Schlegel et al. (1998). 
To obtain an independent determination of the color excess in $(B-V)$ color index, the value of $E(H-K_{\mathrm{S}})$ was transformed into $E(B-V)$ applying the relation $\frac{E(H-K_{\mathrm{S}})}{E(B-V)} = 0.209$ taken from Schlegel et al. (1998). We derived $E'(B-V)=1.24$.
The results seem to be consistent with a mean value of $\left\langle {E}(B-V) \right\rangle =1.21\pm0.04$.

\subsection{Color-magnitude diagrams}

\begin{figure}
\centering
\includegraphics[width=7cm]{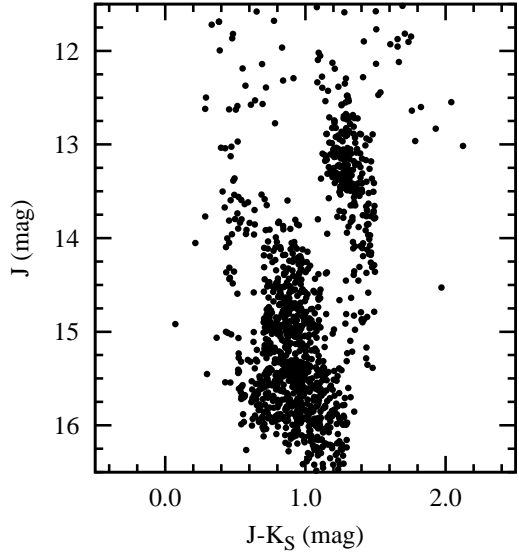}
\caption{The near-IR color-magnitude diagram constructed for the overall ($r<r_{\rm{lim}}$) area of Berkeley~53, after running the decontamination procedure. A morphology typical for a rich old stellar cluster is clearly visible.}
\label{rys4}
\end{figure}

A preliminary $J$ $vs.$ $(J-K_{\mathrm{S}})$ color-magnitude diagram (CMD) for overall cluster region ($r<r_{\mathrm{lim}}$) is presented in Fig.~\ref{rys4}. We applied a decontamination procedure to remove background-star contamination. The details of this procedure can be found in Maciejewski \& Niedzielski (2007). 
The CMDs were built for the cluster region and for an offset field. A concentric offset field of width $10'$ and starting at $r=r_{\rm{lim}}+2'$ from the cluster center was used. Then the CMDs were divided into two-dimensional bins and the number of stars within each box was counted. The cleaned (decontaminated) cluster CMD was built by subtracting the number of stars in the offset box from the number of stars in the corresponding cluster box. The latter number was weighted by the cluster to offset field area ratio. Knowing the number of cluster stars occupying each box, the algorithm randomly chose the required number of stars located in the cluster area and with the adequate magnitude and color index.

As one can see in Fig.~\ref{rys4}, a rich main sequence is clearly visible, as well as a red giant clump. The estimated number of observed cluster stars is $\sim$1000 after rejecting stars with outstanding magnitudes and colors. This value is comparable with the one obtained from the RDP analysis. We found $\sim$200 evolved stars constituting the red clump.

\begin{figure*}
\centering
\includegraphics[width=14cm]{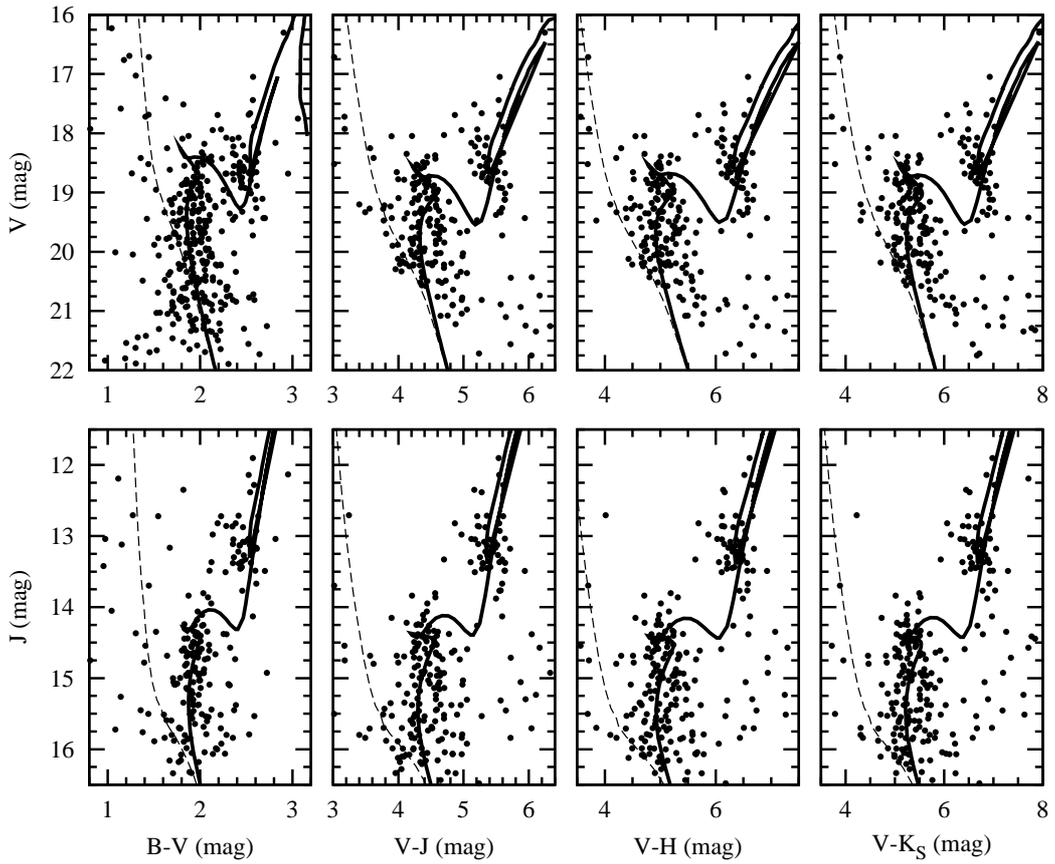}
\caption{The color-magnitude diagrams constructed for the central ($r<r_{\rm{c}}$) part of Berkeley~53. The best-fit isochrone is drawn with a solid line and the zero age main sequence is sketched with a dashed one.}
\label{rys5}
\end{figure*}

To determine fundamental astrophysical parameters of the cluster, we built 8 CMDs combining $V$ and $J$ magnitudes with $(B-V)$, $(V-J)$, $(V-H)$, and $(V-K_{\mathrm{S}})$ color indices. The broad color baselines were expected to minimize photometric errors and to determine precisely the reddening. Only stars located within the core radius around the redetermined cluster center were considered to minimize the influence of the background star contamination on further analysis. The diagrams are plotted in Fig.~\ref{rys5}. 

Initially, we tried to derive the distance modulus and the age of the cluster via fitting a set of theoretical Padova isochrones (Giraldi at al. 2002). Since there is no information about the cluster metallicity, the solar metallicity of $Z=0.019$ was assumed. The fitting algorithm is based on the least-squares method and uses stellar magnitudes as weights. Stars with extremal color indices or magnitudes were rejected manually before running the isochrone-fitting procedure. The fitting and decontamination procedures were run independently for each diagram with a fixed value of reddening. The latter quantity was calculated for a given color using $\left\langle {E}(B-V) \right\rangle$ from the two-color diagram analysis and assuming the universal interstellar extinction law by Schlegel et al. (1998). However, we were unable to find a convergent and satisfactory final solution and the reddening was found to be underestimated.

To explain this discrepancy, we ran the isochrone-fitting algorithm with reddening as a third free parameter. The results of individual fits are collected in Table~1 and the respective isochrones are drawn with solid lines in Fig.~5. The dereddened distance modulus $(m-M)_{0}$ was calculated assuming the universal interstellar extinction law by Schlegel et al. (1998) with a total-to-selective absorption ratio of $R=3.315$. The value of $E(B-V)$ was found to be about 0.3 mag greater than the two-color diagram analysis suggested. One can note that $E(B-V)$ tends to be slightly smaller for colors including more reddish bands. We obtained $\frac{E(V-J)}{E(B-V)}=2.414\pm0.025$, $\frac{E(V-H)}{E(B-V)}=2.707\pm0.017$, and $\frac{E(V-K_{\rm{S}})}{E(B-V)}=2.893\pm0.023$. Comparing to the values given by Schlegel et al. (1998), i.e. 2.413, 2.739, and 2.948, respectively, a discrepancy can be noted in bands $H$ and $K_{\rm{S}}$ and increasing with wavelength. This effect explains the lower value of the interstellar reddening obtained from the TCDs analysis. Adopting redetermined reddening vectors, we obtained consistent results of TCDs and CMDs analysis. In the absence of photometric observations in the $U$ band, it is impossible to answer the question about the nature of the observed effect. It may be caused by an abnormal interstellar extinction law toward Berkeley~53 or by systematics in 2MASS photometry.   

\begin{table*}
\centering
\caption{The results of a solar-metallicity isochrone fit for individual CMDs. $E$ denotes the fitted value of the color excess for a given color, $(m-M)$~-- the fitted value of the apparent distance modulus, $E(B-V)$~-- the calculated $E(B-V)$ color excess, and $(m-M)_{0}$~-- the calculated dereddened distance modulus.} 
\label{tabela1}
\begin{tabular}{l c c c c c}
\hline
Diagram & $\log(age)$ & $E$  & $(m-M)$ & $E(B-V)$ & $(m-M)_0$ \\
        &             &(mag) & (mag)   & (mag)    &  (mag)    \\
\hline 
$V~vs.~(B-V)$          & $9.05$ & $1.54$ & $17.40$ & $1.54$ &$12.35$  \\
$V~vs.~(V-J)$          & $9.10$ & $3.72$ & $17.53$ & $1.54$ &$12.48$  \\
$V~vs.~(V-H)$          & $9.10$ & $4.15$ & $17.49$ & $1.52$ &$12.44$  \\
$V~vs.~(V-K_{\rm{S}})$ & $9.10$ & $4.45$ & $17.52$ & $1.51$ &$12.47$  \\
$J~vs.~(B-V)$          & $9.10$ & $1.53$ & $13.76$ & $1.53$ &$12.39$  \\
$J~vs.~(V-J)$          & $9.10$ & $3.69$ & $13.84$ & $1.53$ &$12.47$  \\
$J~vs.~(V-H)$          & $9.10$ & $4.16$ & $13.88$ & $1.52$ &$12.51$  \\
$J~vs.~(V-K_{\rm{S}})$ & $9.10$ & $4.43$ & $13.87$ & $1.50$ &$12.50$  \\
\hline
\end{tabular}
\end{table*}
 
Finally, the following mean results were obtained from the CDMs analysis: $\log(age)=9.09\pm0.02$, $E(B-V)=1.52\pm0.01$, $(m-M)_0=12.45\pm0.05$, and a distance of $3.1\pm0.1$ kpc. The linear diameter was found to be $19.8\pm0.5$ pc. The main possible cause of systematic error in the values given above is the unknown metallicity of the cluster. To estimate its influence on the results, we repeated the isochrone-fitting procedure for the super- and sub-solar metallicities of $Z=0.030$, $Z=0.008$, and $Z=0.004$. We noticed the increase of the reddening with the decrease of the metallicity ($\Delta E(B-V) \approx 0.07$ for each metallicity step, including $Z=0.019$) while the age and the apparent distance modulus remained stable. It is also worth noting that the discrepancy between the TCDs and CMDs values of the reddening cannot be justified by non-solar metallicity of the cluster.

\subsection{Total mass and number of members}

Studies of the mass function (MF) were carried out to estimate the total mass of the cluster and the number of its members. The analysis is based on the 2MASS photometry due to the wide field of view available and uses the algorithm adopted from Maciejewski \& Niedzielski (2007). The first step was to build a luminosity function (LF) for overall ($r<r_{\rm{lim}}$) cluster region. The bright end of the LF was determined by the main-sequence turn-off point while the faint end was set for $J=15.8$ mag~-- the value of the 99.9\% Point Source Catalogue Completeness Limit\footnote{Following the Level 1 Requirement, according to \textit{Explanatory Supplement to the 2MASS All Sky Data Release and Extended Mission Products} (http://www.ipac.caltech.edu/2mass/releases/allsky/doc)}. We used bins as small as 0.1 mag due to cluster richness. Another LF was built for a concentric offset field of width $10'$ and starting at $r=r_{\rm{lim}}+2'$. The LF of the offset field was subtracted, bin by bin, from the cluster LF, taking the area proportion into account. The resulting LF was converted into an MF using the respective isochrone. It is woth noting that the cluster area, especially the core ($r<r_{\rm{c}}$), may suffer significant incompletness due to stellar crowding. Therefore, our further results must be treated as a lower limit. 

\begin{figure}
\centering
\includegraphics[width=9.0cm]{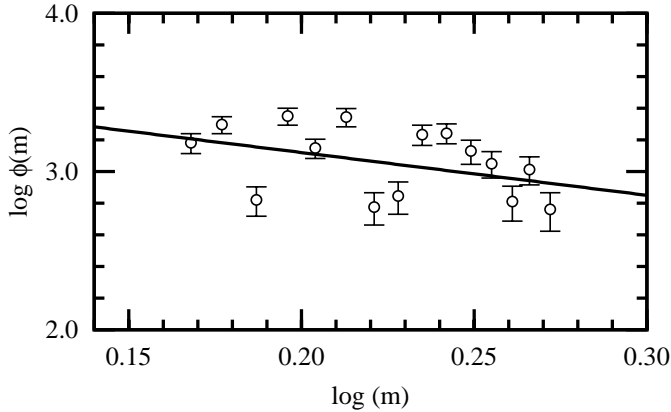}
\caption{The mass function derived for the main-sequence stars of Berkeley~53. }
\label{rys6}
\end{figure}

The mass function $\phi(m)$ was approximated by a standard relation of the form $\log\phi(m)=-(1+\chi)\log{m}+b_{0}$ where $m$ is the stellar mass, $\chi$ is the mass-function slope, and $b_{0}$ is a constant. The derived parameters allowed us to estimate the total mass $M_{\rm{tot}}$ and the total number of stars $N_{\rm{tot}}$. These quantities were calculated extrapolating the MF from the main-sequence turn-off point down to the H-burning mass limit of 0.08 $M_{\odot}$ (see Maciejewski \& Niedzielski 2007 and references therein for details). The mass function is plotted in Fig.~\ref{rys6}. We obtained $\chi=1.7\pm1.6$, a value comparable within error bars to the universal initial mass function (IMF) $\chi_{\rm{IMF}}=1.3 \pm 0.3$ given by Kroupa (2001). The high uncertainty is caused by a small range of covered stellar masses. The cluster was found to be in fact very rich and massive with $N_{\rm{tot}}=31000$ and $M_{\rm{tot}}=12000$ $\rm{M}_{\odot}$.

To describe a state of cluster dynamic evolution, the dynamical-evolution parameter $\tau$ was calculated in the form $\tau=\frac{age}{t_{\rm{relax}}}$ where $t_{\rm{relax}}$ is the relaxation time (see Maciejewski \& Niedzielski 2007 for details). We derived $\log\tau=-0.3$ which suggests that the cluster is dynamically younger than its relaxation time.

\section{Summary}

Berkeley~53 was found to be a rich and massive open cluster belonging to the Perseus Arm. Its age exceeds 1 Gy, but it seems to be very young in the context of its dynamical evolution. A subtle discrepancy in the interstellar extinction law toward the cluster in the $H$ and $K_{\rm{S}}$ bands was detected. That suggests the presence of an abnormal interstellar extinction law toward the cluster.

\textit{Acknowledgements}: 
We thank the referee for remarks that improved our paper. This research has made use of the WEBDA and SIMBAD data bases and is supported by UMK grant 411-A and the grant VU-NZ-01/06 of the Ministry of Education and Science of Bulgaria. GM acknowledge support from the EU in the FP6MC ToK project MTKD-CT-2006-042514. This publication makes use of data products from the Two Micron All Sky Survey, which is a joint project of the University of Massachusetts and the Infrared Processing and Analysis Center/California Institute of Technology, funded by the National Aeronautics and Space Administration and the National Science Foundation.

\end{document}